\documentclass[prl,twocolumn,showpacs,preprintnumbers,floatfix]{revtex4}
\usepackage{textcomp}
\usepackage{amsmath}
\usepackage{amssymb}
\usepackage{graphicx}

\begin{document}

\title{Ultra-high-$Q$ tunable whispering-gallery-mode microresonator}
\author{M. P\"ollinger, D. O'Shea, F. Warken}

\affiliation{Institut f\"ur Physik, Johannes Gutenberg-Universit\"at Mainz, 55099 Mainz, Germany}

\author{A. Rauschenbeutel}

\email{rauschenbeutel@uni-mainz.de}

\affiliation{Institut f\"ur Physik, Johannes Gutenberg-Universit\"at Mainz, 55099 Mainz, Germany}

\date{\today}

\begin{abstract}
Typical microresonators exhibit a large frequency spacing between
resonances and a limited tunability. This impedes their use in a
large class of applications which require a resonance of the
microcavity to coincide with a predetermined frequency. Here, we
experimentally overcome this limitation with highly prolate-shaped
whispering-gallery-mode ``bottle microresonators'' fabricated from
standard optical glass fibers. Our resonators combine an ultra-high
quality factor of 360 million, a small mode volume, and near
lossless fibre coupling, characteristic of whispering-gallery-mode
resonators, with a simple and customizable mode structure enabling
full tunability.
\end{abstract}


\maketitle

Optical microresonators hold great potential for many fields of
research and technology \cite{Vahala03}. They  are used for filters
and switches in optical communications \cite{Djordjev02,Chu99,
Almeida04}, non-linear optics \cite{Del'Haye07}, bio(chemical)
sensing \cite{Armani07}, microlasers
\cite{Sandoghdar96,vonKlitzing00,Cai00a}, as well as for cavity
quantum electrodynamics applications such as  single photon sources
\cite{Michler00,McKeever04,Hijlkema07} and interfaces for quantum
communication \cite{Boozer07,Wilk07}. All these applications rely
on the spatial and temporal confinement of light by the
microresonator, characterized by its mode volume $V$ and its
quality factor $Q$, respectively \cite{Vahala03}. The ratio $Q/V$
thus defines a key figure relating the coupling strength between
light and matter in the resonator to the dissipation rates of the
coupled system. The highest values of $Q/V$ to date have been
reached with whispering-gallery-mode (WGM) microresonators
\cite{Kippenberg04}. Standard WGM microresonators, like dielectric
microspheres, microdisks, and microtori, typically confine the
light in a narrow ring along the equator of the structure by
continuous total internal reflection at the resonator surface
\cite{Matsko06}. While such equatorial WGMs have the advantage of a
small mode volume they also exhibit a large frequency spacing
between consecutive modes. In conjunction with the limited tuning
range due to their monolithic design, tuning of equatorial WGM
microresonators to an arbitrary frequency has therefore not been
realized to date.

\begin{figure}
 \centering
 \includegraphics[width=0.285\textwidth]{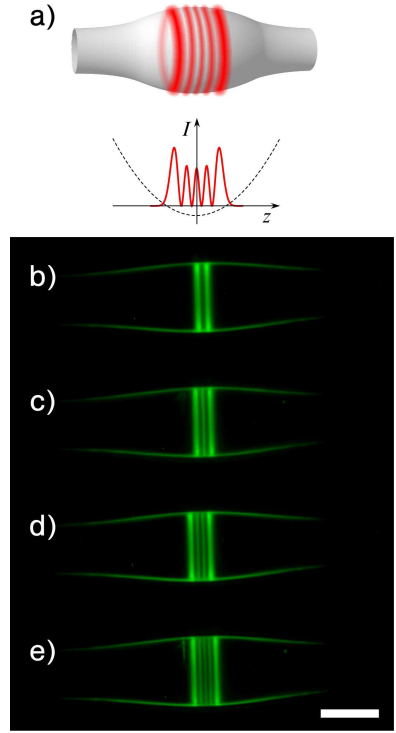}
    \caption{\label{fig1} (a) Concept of the bottle microresonator. In
    addition to the radial confinement by continuous total internal
    reflection at the resonator surface, the axial confinement of the
    light is caused by an effective harmonic potential (dashed line)
    fixed by the curvature of the resonator profile. The resulting
    intensity distribution is therefore given by the eigenfunctions of
    the quantum mechanical harmonic oscillator \cite{Louyer05}.
    (b)--(e) Experimental micrographs of the $q=1$--4 bottle modes visualised via
    the upconverted green fluorescence of dopant erbium ions in a 36-$\mu$m
    diameter bottle microresonator. Scale bar, 30 $\mu$m.}
\end{figure}

For this reason, the WGM ``bottle microresonator'' has recently
received considerable attention
\cite{Kakarantzas01,Ward06,Warken08,Strelow08} because it promises
a customizable mode structure while maintaining a favourable $Q/V$
ratio \cite{Sumetsky04,Louyer05}. Due to its highly prolate shape,
the bottle microresonator gives rise to a class of
whispering-gallery-modes (WGMs) with advantageous properties, see
Fig.~\ref{fig1}(a).
The light in these ``bottle modes'' harmonically oscillates back
and forth along the resonator axis between two turning points which
are defined by an angular momentum barrier \cite{Louyer05}. The
resulting axial standing wave structure exhibits a significantly
enhanced intensity at the so-called ``caustics'' of the bottle
mode, located at the turning points of the harmonic motion. The
bottle microresonator possesses an equidistant spectrum of
eigenmodes, labelled by the ``azimuthal quantum number'' $m$, which
counts the number of wavelengths that fit into the circumference of
the resonator, and the ``axial quantum number'' $q$, which counts
the number of axial intensity nodes \cite{Sumetsky04,Louyer05}. The
frequency spacing between modes with consecutive quantum numbers
$q$ ($m$) is called the axial (azimuthal) free spectral range and
will be denoted $\Delta\nu_q$ ($\Delta\nu_m$) in the following.

In small WGM resonators the azimuthal free spectral range (FSR) is
typically very large. For example, changing $m$ by one for a
35-$\mu$m diameter WGM changes its resonance frequency by
$\Delta\nu_m\approx 2$ THz, i.e., about one percent of the optical
frequency. Tuning a WGM microresonator over such a large range is a
critical issue. Exploiting the temperature dependence of the
refractive index of silica, electrical thermo-optic tuning of
equatorial WGMs in a 75-$\mu$m diameter microtorus over more than
300 GHz, i.e., up to 35 \% of the azimuthal FSR, has been
demonstrated \cite{Armani04}. Another tuning scheme involves
elastically deforming the resonator through mechanical strain,
thereby changing its diameter and the refractive index of the
medium. Using this strain tuning technique, tuning over 400 GHz
(50~\% of the azimuthal FSR) has been demonstrated for a 80-$\mu$m
diameter microsphere, limited by the mechanical damage threshold of
the resonator \cite{vonKlitzing01}. The bottle microresonator
circumvents this problem: Its axial FSR only depends on the
curvature of the resonator profile and can thus be made
significantly smaller than its azimuthal FSR \cite{Louyer05}. It is
therefore sufficient to tune the bottle microresonator over one
axial FSR in order to ensure that an arbitrary predetermined
frequency will coincide with the resonance frequency of an
appropriately chosen bottle mode.

Previous experimental work has demonstrated that bottle
microresonators are readily fabricated from standard optical glass
fibers with an initial diameter of 125~$\mu$m in a two-step
heat-and-pull process \cite{Kakarantzas01,Ward06,Warken08}: First,
a few millimetres long section with a homogeneous diameter is
created by elongating the fiber while heating the section with a
travelling flame or a scanned CO$_2$-laser beam. Next, a bulge is
created on the tapered fiber waist which is located between two
microtapers. Each of the microtapers is realized by locally heating
the tapered fiber waist with a focussed CO$_2$-laser beam while
slightly stretching it. The resulting bulge forms the bottle
microresonator which exhibits a parabolic variation of the fiber
diameter $D(z)\approx D_0\cdot[1-(\Delta k\cdot z)^2/2]$ around its
central zone, where $D_0$ is the maximum diameter of the bulge at
the position $z = 0$ along the resonator axis and $\Delta k$
denotes the curvature of the resonator profile. So far, the values
for $D_0$ ranged from 12~$\mu$m \cite{Ward06} to 16~$\mu$m
\cite{Kakarantzas01,Warken08}. Theoretically, these values should
be large enough to avoid radiative losses and thus to reach $Q$
factors in the $10^7$--$10^9$ range \cite{Buck03}. However, the
experimentally observed $Q$ factors were smaller than $10^4$ in
\cite{Kakarantzas01}. In our own work, we observed a strong
decrease of the $Q$ factor for diameters below $D_0\approx
30~\mu$m, in agreement with the results obtained on microtori
\cite{Kippenberg04}. Therefore, the bottle microresonators used
here have a diameter of $D_0= 30$--40~$\mu$m. We reconstruct the
diameter profile with a precision of $\pm 2$~$\mu$m using a
microscope in combination with a customized image analysis
software. This method allows us to determine $\Delta k$ with a
precision of $\pm 0.001$~$\mu$m$^{-1}$. Adjustment of the
CO$_2$-laser beam spotsize, the microtaper separation, and the
elongation length allows us to precisely tailor the resonator shape
and thereby customize its mode structure. Typically, we work with a
curvature of $\Delta k = 0.010$--0.015~$\mu$m$^{-1}$.

The spatial and spectral properties of the modes are investigated
using a distributed feedback (DFB) diode laser operating around 850
nm with a short-term ($< 5$~$\mu$s) linewidth of 400 kHz. Efficient
coupling of propagating light fields to the bottle modes requires
phase matched excitation. We couple light in and out by means of a
sub-micron sized tapered ``coupling fiber'' \cite{Knight97}. It is
fabricated by a similar heat-and-pull technique as described above
and is aligned perpendicularly to the resonator axis at one of the
caustics of a particular axial mode. Its radius is chosen for
optimized phase matching to the lowest order radial modes which
exhibit the smallest mode volume and the highest evanescent field
at the surface of the resonator \cite{Knight97}. The gap between
the coupling fiber and the resonator is comparable to the decay
lengths of the evanescent fields of both structures, i.e., a few
hundred nanometres, and is controlled with a resolution of 10 nm
using a piezoelectric actuator. For coupling to different axial
modes the position of the coupling fiber can be scanned along the
resonator axis via a servomotor-driven translator with 100 nm
resolution. By carefully adjusting the coupling fiber-resonator
gap, we realize so-called critical coupling where the incident
optical power is entirely dissipated in the resonator and the
transmission of the coupling fiber at resonance ideally drops to
zero \cite{Cai00b}. In addition, we experimentally observe
near-unity values ($> 99.5$~\%) for the ideality \cite{Spillane03}
of our fiber taper coupler. This means that the coupling junction
between the fiber taper and the bottle microresonator introduces
only very weak losses. Unity ideality holds huge importance for a
wide range of applications where near-lossless transfer of light to
and from the resonator mode is a requisite.

In order to characterize the spatial properties of the bottle modes
we fabricate an erbium doped resonator with $D_0 = 36$~$\mu$m and
$\Delta k = 0.015$~$\mu$m$^{-1}$ from the 50-$\mu$m diameter core
of a standard Er$^{3+}$-doped multimode fiber, whose cladding is
removed by wet etching. When resonantly exciting bottle modes
around 850 nm, the erbium ions emit fluorescence light around 540
nm in an upconversion process \cite{vonKlitzing00,Cai00a}. This
green fluorescence is then observed using an optical microscope.
Depending on the position of the coupling fiber along the resonator
axis and the laser frequency different axial modes can be
individually excited, see Fig.~\ref{fig1}~(b)--(e). Each picture
has been generated from a stack of micrographs obtained by varying
the focal plane in order to increase the effective focal depth. The
measured intensity distribution confirms the axial standing wave
structure and the enhanced light intensity at the caustics which
are at the heart of the bottle microresonator concept
\cite{Louyer05,Strelow08}.

\begin{figure}
 \centering
 \includegraphics[width=0.4\textwidth]{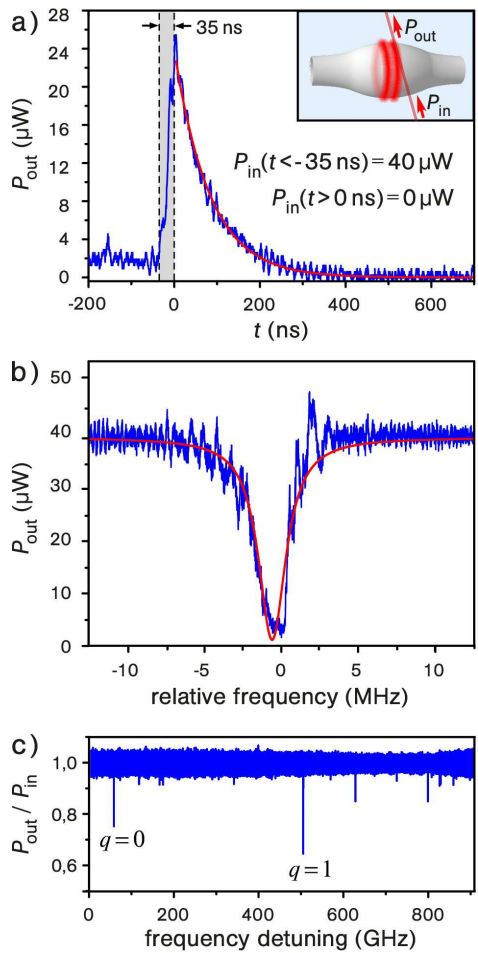}
    \caption{\label{fig3} (a) Cavity ringdown measurement of the $q = 1$ bottle
    mode in a bottle microresonator with 35 $\mu$m diameter and a
    curvature of $\Delta k = 0.012$~$\mu$m$^{-1}$ at critical coupling
    for a wavelength near 850 nm and transverse electric (TE)
    polarization, i.e., with the magnetic field parallel to the
    resonator surface. The inset
    schematically shows the coupling geometry with the sub-micron
    coupling fiber aligned with one of the two caustics of the bottle
    mode. (b) Spectrum of a 38~$\mu$m-diameter
    bottle microresonator with  $\Delta k = 0.009$~$\mu$m$^{-1}$, obtained by scanning
    the probe laser frequency over
    the resonance of a $q\approx 30$ bottle mode at critical coupling. The solid line is a
    Lorentzian fit yielding a FWHM linewidth of $2.1\pm 0.1$~MHz, corresponding to
    an intrinsic quality factor in excess of $Q_0 = 3.3 \times 10^8$. Due to the
    presence of optical bistability the measured line is slightly asymmetric
    and broadened. (c) Wide range transmission spectrum showing both the  $q = 0$ and
    $q = 1$ bottle mode in a 37~$\mu$m-diameter
    bottle microresonator with $\Delta k = 0.012$~$\mu$m$^{-1}$.
    }
\end{figure}

Next we determine the quality factor $Q=\omega\tau$ of an undoped
resonator, where $\tau$ is the photon lifetime in the resonator and
$\omega=2\pi\nu$ is the optical frequency. For this purpose, we use
a cavity ringdown technique: We resonantly excite the bottle mode
under investigation with the 850 nm probe laser at critical
coupling. After switching off the probe beam within 35 ns using an
acusto-optical modulator, the exponential decay of the intracavity
power is monitored through the output port of the coupling fiber.
This measurement, shown in Fig.~\ref{fig3}~(a), is taken on a
resonator with $D_0 =  35$~$\mu$m and $\Delta k =
0.012$~$\mu$m$^{-1}$ and yields a photon lifetime at critical
coupling of $\tau_{\rm crit}= 82$~ns. We thus obtain a lower bound
for the intrinsic photon lifetime in the uncoupled resonator of
$\tau_0 = 2\tau_{\rm crit}= 164$~ns \cite{Kippenberg04} and an
intrinsic quality factor in excess of $Q_0 = 3.6 \times 10^8$. This
ultra-high intrinsic quality factor is comparable to the values
reported for other WGM microresonators of the same diameter
\cite{Kippenberg04}. Spectral measurements, which are, however,
affected by thermal bistability, confirm this result, see
Fig.~\ref{fig3}~(b). We note that at critical coupling, required
for many applications, our quality factor of $Q_{\rm crit}=
\omega\tau_{\rm crit} = 1.8 \times 10^8$ is about one order of
magnitude larger than what has previously been reported
\cite{Kippenberg04}. Several measurements on different resonators
showed similar quality factors, independent of the axial quantum
number. Figure~\ref{fig3}~(c) shows a wide range transmission
spectrum of a bottle microresonator. The coupling to the $q = 0$
and $q = 1$ bottle modes was selectively maximized by optimizing
the phase-matching, resulting in an advantageously low spectral
mode density. The $q$ quantum numbers of the modes were identified
in a separate measurement by translating the coupling fiber along
the resonator axis while recording the spatial modulation of the
coupling efficiency.

\begin{figure}
 \centering
 \includegraphics[width=0.35\textwidth]{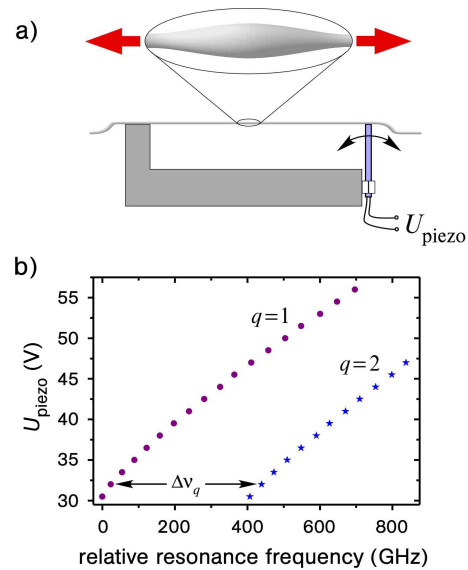}
    \caption{\label{fig4} (a)
    Design of the mount for applying mechanical strain
    to the bottle microresonator. The fiber that contains the
    resonator structure is attached to a piezo-electric bending
    actuator which elastically elongates the fiber and is
    controlled by a voltage $U_{\rm piezo}$. (b) Strain tuning of the TE
    polarized $q = 1$ and $q = 2$ bottle modes of the same bottle
    microresonator as used in Fig.~\ref{fig3}~(a). The tuning range of 700 GHz of the
    $q = 1$ mode exceeds the observed axial FSR of $\Delta\nu_q = 425
    \pm 8$~GHz by a factor of 1.7.}
\end{figure}

We now demonstrate strain tuning of a bottle microresonator using
the setup depicted in Fig.~\ref{fig4}~(a). The measurement was
carried out with the 35-$\mu$m diameter bottle microresonator which
yielded the ultra-high $Q$ factor in the above ringdown
measurement.  In Fig.~\ref{fig4}~(b) the resonance frequencies of
the $q = 1$ and $q = 2$ bottle modes were measured for varying
mechanical strain. The so determined tuning range of 700 GHz
corresponds to 700,000 linewidths of the resonator and is 1.7 times
larger than the axial FSR. The experimentally measured axial FSR of
$\Delta\nu_q = 425 \pm 8$~GHz is in good agreement with the
theoretical value of $397 \pm 33$~GHz, calculated from the measured
curvature of the resonator profile. The maximum stress applied to
the resonator in this measurement, limited by the travel range of
the bending actuator, can be inferred from the frequency shift and
was about 35~\% of the typical damage threshold of the silica
resonator structure \cite{Glaesmann91}. The bottle microresonator
can thus be tuned to an arbitrary frequency by strain tuning the
nearest axial bottle mode. However, in order to keep the mode
volume small it is desirable to work with a low axial quantum
number $q$, thereby minimizing the separation between the two
caustics, see Fig.~\ref{fig1}(b)--(e). Remarkably, this is possible
for the bottle microresonator which exhibits two FSRs, axial and
azimuthal, and which can thus be tuned to any arbitrary frequency
using, e.g., only the $q = 1$--4 axial bottle modes which allow to
bridge the azimuthal FSR. These four lowest order bottle modes have
a calculated mode volume ranging from 1180~$\mu$m$^3$ to
1470~$\mu$m$^3$ for a wavelength of $\lambda\approx 850$~nm,
corresponding to 6070 $(\lambda/n) ^3$ to 7560  $(\lambda/n) ^3$,
where $n = 1.467$ is the index of refraction of silica. This
results in an intrinsic $Q / V$ value of up to $5.9\times 10^4
(\lambda/n) ^{-3}$ which remains as high as $3.0\times 10^4
(\lambda/n) ^{-3}$ in the case of critical coupling. These numbers
are high enough to realize light-matter interaction deep within the
so-called strong coupling regime. For example, the atom-field
coupling rate between a cesium atom and the $q = 1$ bottle mode
near the surface of the resonator is calculated to be $g =
2\pi\times 50$~MHz which clearly exceeds the dissipation rates of
the system, given by the cavity field decay rate
$\kappa=\omega/2Q_0 = 2\pi\times 0.49$~MHz and the transverse
atomic dipole decay rate $\gamma_\perp = 2\pi\times 2.6$~MHz.

Summarising, we present a fully tunable whispering gallery mode
microresonator. The tunability of our bottle microresonator stems
from the confinement of the light between two caustics in a simple
axial mode structure. In conjunction with its high $Q / V$ value
this reveals the enormous potential of the bottle microresonator
for coupling light and matter. Moreover, the advantageous mode
geometry of the bottle modes allows near-lossless simultaneous
coupling of two independent coupling fibers at the two caustics and
thus to induce a resonator-mediated non-linear interaction between
two distinct optical signals at the single photon level. The bottle
microresonator thereby opens the route towards the realisation of
next-generation communication and information processing devices
such as single photon all-optical switches \cite{Bermel06} and
single photon transistors \cite{Chang07}.

The authors wish to thank W. Alt and D. Meschede for their valuable
support and in-depth discussions. Financial support by the DFG
(Research Unit 557), the Volkswagen Foundation (Lichtenberg
Professorship), and the ESF (EURYI Award) is gratefully
acknowledged.

\end{document}